# Charged Black Holes and Constraints on Baryon Asymmetry

C Sivaram and Kenath Arun

Indian Institute of Astrophysics

**Abstract:** The no-hair theorem, which postulates that all black holes can be completely characterized by only three externally observable parameters: mass, electric charge, and angular momentum, sets constraints on both the maximal angular momentum and maximal electric charge. In this work, we would explore the consequence of these for the formation of primordial black holes in the early universe and also the formation of black holes due to collapse of dark matter configurations and how this could be used to probe the conditions in the very early universe and constrain the epoch when baryon asymmetry was established.

Black holes irrespective of how and where they form are characterised [1] by only three conserved parameters (1) Total mass, (2) Net electric charge and (3) Angular momentum. To preserve the event horizon, the maximal angular momentum and maximal electric charge are well known to be constrained by the relations: [2, 3]

$$J_{max} \leq \frac{GM^2}{c} \qquad \ldots (1)$$

$$Q_{max} \leq \sqrt{G}M \qquad \ldots (2)$$

$Q_{max}$ is the net maximal electric charge and $J_{max}$ is the maximal angular momentum allowing formation of the black hole.

Here we would explore the consequence of equation (2), for the formation of primordial black holes (PBH) in the early universe and also the formation of black holes due to collapse of dark matter (DM) configurations.

To begin with, we note that for solar mass black hole, the net maximal electric charge is:

$$Q_{max} = \sqrt{G}M_{sun} \approx 5 \times 10^{29} \, esu \approx 10^{39} e \qquad \ldots (3)$$

Where $e$ is the electron charge

As the total number of nucleons which has undergone collapse is $\sim 10^{57}$, this implies that there is less than one part in $10^{19}$ excess charge per nucleon (i.e. the number of positive and negative charge must be balanced to one part in $10^{19}$).

Typically for a neutron star, to avoid n decay, the electron (and proton) number density is typically $\sim 10^{-3}$ of the neutron density. If a 10 solar mass black hole is formed, the charges of opposite sign should be equal to one part in $10^{20}$! (This approaches the experimental limit [4] on equality of charge on proton and electron, i.e. $\sim 10^{-22}$). So in case of black holes forming by accretion, the charge neutrality of the accreted matter (plasma under the usual astrophysical situations) must be balanced to this order! A higher ratio would prevent formation of the black hole.

For a supermassive black hole, this constraint is even more stringent [5]. For $M_{BH} \approx 10^9 M_{sun}$, $Q_{max} < 10^{38} esu$ or the fractional excess allowed is $< 10^{-20}$.

This charge imbalance should also constrain [6] PBH's formed in the early universe. in the early universe [7, 8] primordial black holes form when the metric fluctuations exceed unity and for example in the radiation pressure forces material inside the Schwarzschild radius provided it began with a density in excess of the ambient density. Using the relation for the energy density in the radiation era as a function of time given by:

$$\rho_R = \frac{3}{32\pi G t^2} \quad \ldots (4)$$

The total mass of the radiation energy in causal contact after time $t$ is given by:

$$M_H = \frac{4\pi}{3} c^3 t^3 \rho_R \quad \ldots (5)$$

And using equation (4) for $\rho_R$ this implies that at an epoch $T < t$, only black holes of mass give by the following can form [9].

$$M_{BH} = \frac{c^3 t}{8G} \quad \ldots (6)$$

The evaporation time for a PBH, due to Hawking radiation is given by: [9, 10]

$$t_{ev} = \frac{5120\pi G^2 M_{BH}^3}{\hbar c^4} \qquad \ldots (7)$$

Thus the mass of the PBH with evaporation time comparable to the Hubble age (H) is:

$$M_{BH}^{(H)} \approx 10^{14} g \qquad \ldots (8)$$

The corresponding epoch in the early universe at which such black holes can form is given from equation (6) as:

$$t_{BH}^{(H)} \approx \frac{8GM_{BH}^{(H)}}{c^3} \approx 2\times 10^{-24} s \qquad \ldots (9)$$

The corresponding temperature of the radiation era at this epoch is then:

$$T = \frac{10^{10} K}{t_{(s)}^{1/2}} \approx 10^{22} K$$

This corresponds to an energy which is $\sim 10^9$ the proton rest energy $\sim 10^{-15} g$. So if at this epoch the universe has equal numbers [11] of massive particles having baryon number and charge of both signs, i.e. equal numbers of baryons and antibaryons, the number of particles requires to form such a PBH, whose mass is given by equation (8) is:

$$N \approx \frac{10^{14} g}{10^{-15} g} \approx 10^{29}$$

For a $10^{14} g$ PBH, maximal charge is:

$$Q_{max} = \frac{\sqrt{G}M}{e} \approx 10^{20} \qquad \ldots (10)$$

So the constraint on the asymmetry between charges (and by implication baryon number) for formation of the PBH's (evaporating at the current epoch) is $\Delta Q \approx \Delta B < 10^{-9}$, or one part in $10^9$. So this clearly implies that even at $t \approx 10^{-24} s$ in the early universe, the baryon asymmetry (in this case, the baryon number and charge carried by massive particles and antiparticles with energies corresponding to that epoch) should already be at the level of 1 in $10^9$, otherwise the PBH's would not form and would not be evaporating now.

In principle PBH's evaporating at present epoch, could probe the conditions in the very early universe (at $>10^7 TeV$) and constrain the epoch when baryon asymmetry was established.

Similar observation of heavier mass PBH's (by other techniques) could provide information about the baryon asymmetry at later epochs.

Again DM particles (with mass range from 10GeV to several TeV) could collapse to form black holes [12, 13]. The typical mass of such a black hole is given by:

$$M_{BH}^{(DM)} = \frac{m_{pl}^3}{m_D^2} \quad \ldots (11)$$

Where $m_{pl}$ is the Planck mass and $m_D$ is the DM particle mass.

The maximal electric charge is given by: $Q_{max} = \frac{\sqrt{G}}{e} \frac{m_{pl}^3}{m_D^2}$

This gives the constraint on the charge asymmetry (assuming DM particles could carry charge of both signs) of

$$Q \leq \frac{G}{e} m_D \quad \ldots (12)$$

For a 1MeV DM particle this is $Q < 10^{-21} e$

This implies that even in the formation of DM particles in the early universe, there could be asymmetries of this order in their initial production. This asymmetry would be preserved at later epochs. The LHC in principle could test such possibilities [14].